\documentclass[11pt,a4paper]{article}
\pdfoutput=1
\usepackage{graphicx}
\usepackage{physics}
\usepackage{multirow}
\usepackage[T1]{fontenc}
\usepackage[utf8]{inputenc}
\usepackage{authblk}
\usepackage{hyperref}

\title{Neutrino phenomenology from leptogenesis}

\author[1]{Franco Buccella\thanks{buccella@na.infn.it }}
\author[1,2]{Damiano F. G. Fiorillo\thanks{damianofg@gmail.com}}
\author[1,2]{Gennaro Miele\thanks{miele@na.infn.it}}
\author[1,2]{Stefano Morisi \thanks{stefano.morisi@gmail.com}}
\author[1,2]{Ofelia Pisanti\thanks{pisanti@na.infn.it}}
\author[1,2]{Pietro Santorelli\thanks{Pietro.Santorelli@na.infn.it}}
\affil[1]{{\it INFN - Sezione di Napoli, Complesso Univ. Monte S. Angelo, I-80126 Napoli, Italy}}
\affil[2]{{\it Dipartimento di Fisica {\it "Ettore Pancini"}, Universit\`a degli studi di Napoli "Federico II", Complesso Univ. Monte S. Angelo, I-80126 Napoli, Italy}}

\begin{document}

\maketitle

\begin{abstract}
Assuming a type-I seesaw mechanism for neutrino mass generation and invoking a baryogenesis \textit{via} leptogenesis scenario, we consider a reasonable hierarchical structure for Dirac neutrino mass matrix, similar to up-type quark mass matrix. These hypotheses imply a relevant connection between  high scale CP violation and low energy one. By requiring a compact  heavy neutrino mass spectrum, which allows to circumvent Davidson-Ibarra limit, one can obtain an efficient leptogenesis restricting the allowed region for low energy neutrino parameters. Once the oscillating parameters are taken inside a $3\sigma$ range, through the numerical resolution of the leptogenesis Boltzmann equations one gets the following allowed intervals for the lightest neutrino mass and the Dirac CP phase: $-0.90\pi<\delta<-0.75\pi$ and   $m_1\sim( 0.002 - 0.004)$\,eV. 
\end{abstract}

\section{Introduction}
As well known, neutrino oscillations imply non vanishing  neutrino masses that require an extension of the electroweak Standard Model (SM). In this scheme one gets for free a possible mechanism to explain the baryon asymmetry of the universe (BAU). In this paper we consider the particular case in which SM is extended by adding three right-handed Standard Model singlets $N_{i}$ with Yukawa interactions
\begin{eqnarray}\label{Yu}
\mathcal{L} =&&-\sum_{\alpha \beta}\left(Y^\ell\right)_{\alpha \beta} \overline{L}_{L \alpha} H\ell_{R\beta} -\sum_{\alpha i}\left(Y^\nu \right)_{\alpha i} \overline{L}_{L \alpha} \tilde{H} N_i +\nonumber\\
&&- \frac 12 \sum_{ij}\left(M_R\right)_{ij} \overline{N^{c}_i} N_j 
\end{eqnarray}
where $\alpha,\beta,i,j=1,2,3$ represent family indexes. In the basis where charged leptons are diagonal the three active light neutrinos get a Majorana mass from the type-I seesaw  mechanism \cite{Minkowski:1977sc,Yanagida:1979as,GellMann:1980vs,Mohapatra:1979ia} that in the family space reads
\begin{eqnarray}\label{ss1}
M_\nu &=& -M_D \left(M_R\right)^{-1}M_D^T\,,
\end{eqnarray}
where the matrix $M_D\equiv Y^\nu \langle H^0 \rangle$, and neutrino mixing parameters and masses are fitted from the observations. In this framework the CP violating out of equilibrium decay of $N_{i}$ can produce a lepton asymmetry that is then converted into a baryon asymmetry by sphalerons. Such a mechanism is known as {\it baryogenesis via leptogenesis} \cite{Fukugita:1986hr} and can provide a viable origin for BAU. The possible correlations between the high energy scale CP violation and the low energy one, which comes from the diagonalization of Eq. (\ref{ss1}) and it is still waiting for an experimental confirmation, has been intensively studied in literature
\cite{Berger:1999bg,Buchmuller:2000as,Akhmedov:2003dg,Pascoli:2006ie,Pascoli:2003uh,Joshipura:2001ui,Falcone:2000ib,Hirsch:2001dg,Buccella:2001tq,Goldberg:1999hp,Mohapatra:2006se,Antusch:2005tu,Ellis:2002eh,Frampton:2002qc}.

The existence of such kind of relation can be naively understood by inverting the relation of Eq.\,(\ref{ss1}), namely 
\begin{eqnarray}\label{mR}
M_R&=&-M_D^T M_\nu^{-1} M_D\,.
\end{eqnarray}
The lepton asymmetry generated by leptogenesis depends on the right handed neutrino couplings and their masses, namely by the matrices $M_R$ and $M_D$. By fixing the Dirac neutrino mass matrix $M_D$, one gets from Eq.\,(\ref{mR}) a connection between light neutrino mass matrix $M_\nu$ and  lepton asymmetry  that is related to the high energy scale CP violation. Unfortunately the experiments do not provide direct information about the Dirac mass matrix, and therefore in order to fix $M_D$ we have to evoke some theoretical arguments as also given for example in \cite{Berger:1999bg}, \cite{Akhmedov:2003dg} and \cite{Frampton:2002qc}. Furthermore, one can observe that from Yukawa interactions of Eq.\,(\ref{Yu}), in order to fit the charged lepton masses, the Yukawa matrix must be very hierarchical, namely  $Y^\ell_{11}\ll Y^\ell_{22} \ll Y^\ell_{33}$. The origin of such hierarchy represents one of the biggest challenge of flavour physics, and several ideas has been developed to solve such a puzzle, like for instance the use of Froggatt-Nielsen (FN) flavour symmetries \cite{Froggatt:1978nt,Altarelli:1999wi}. In this paper we do not focus our attention on such a problem, but rather we observe that, belonging the left-handed neutrinos to the same $SU(2)_L$ doublets of the charged leptons, if some symmetry enforces a hierarchy in the charged lepton Yukawas, it is reasonable  that some hierarchy is also present in the Dirac neutrino coupling $Y^\nu$. Another possibility to fix the structure of the Dirac mass matrix is to assume a Grand Unified gauge group like $SO(10)$  that implies $M_D\approx M_{\textrm{up}}$, where $M_{\textrm{up}}$ denotes the up-type quark mass matrix\footnote{Note that the unrealistic $SO(10)$ model with only one {\bf 10}-Higgs scalar field predicts exactly $M_D= M_{\textrm{up}}$. }. These two examples will be shortly described in appendix, while here we just remark that both scenarios imply a hierarchical Dirac Yukawa matrix, namely almost diagonal and with  $Y^\nu_{11}\ll Y^\nu_{22} \ll Y^\nu_{33}$. It follows that  also the right-handed neutrino mass matrix results to be hierarchical unless the entries of the matrix $M_\nu$ are strongly constrained. Thus, barring any particular assumption we expect a hierarchical heavy right-handed neutrino mass spectrum
\begin{eqnarray}
M_{R_1}\ll M_{R_2} \ll M_{R_3}   \,,
\end{eqnarray}
with $M_{R_1}$, $M_{R_2}$ and $M_{R_3}$ denoting the eigenvalues of $M_R$. Note that if the hierarchy is too strong it could give problem in the origin of  baryogenesis via  leptogenesis. Indeed, in case the heaviest of the right-handed neutrinos has a mass of about $M_{R_3} \approx 10^{14}$ GeV (that is around the  grand unified scale) the lightest right-handed mass could be below the Davidson-Ibarra limit $M_{R_1} <10^9$ GeV \cite{Davidson:2002qv}, which is the lower limit for a lepton asymmetry generated by the decay of the lightest right-handed neutrino to be sufficiently large\footnote{Of course how small is the mass of the lightest right-handed neutrino compared to the heaviest one depends on the specific model considered.}, as shown also in \cite{Akhmedov:2003dg}. 

A mechanism to obtain a reasonable value for the lepton asymmetry in this context has been proposed in \cite{Buccella:2012kc}. The idea is quite simple: by imposing a compact spectrum for the right-handed masses on the l.h.s. of Eq.\,(\ref{mR}), it follows a condition for neutrino mass parameters on the r.h.s. of Eq.\,(\ref{mR}) (see also \cite{Akhmedov:2003dg}) once the structure of the Dirac neutrino mass matrix $M_D$ is fixed (by hand or by some model). In this paper we provide a numerical study of such an approach, and explore possible correlations between neutrino observables coming from the compact right-handed spectrum hypothesis and the prediction for leptogenesis. In the next section we review the notation and we deeper explain the basic idea of \cite{Buccella:2012kc}. In section 2 we review the basic ideas and relations of leptogenesis. In section 4 we report our results with and without imposing the constraint coming from leptogenesis. Then in section 5 we give our conclusions.

\section{Compact right-handed spectrum: implication for neutrino phenomenology}
By following the notation of \cite{Buccella:2012kc} we put
\begin{eqnarray}
M_D=V_L^\dagger \, M_D^{\rm diag} \, V_R
\end{eqnarray}
where $V_{L,R}$ are unitary matrices and $M_D^{\rm diag}=(M_{D_1},M_{D_2},M_{D_3})$. Once the above expression is used in Eq.\,(\ref{mR}) one has 
\begin{eqnarray}
M_R &= &-V_R^T M_D^{\rm diag} \, A \,M_D^{\rm diag} V_R  \,;\label{eqMR}\\
A &\equiv &V_L^* \, M_\nu^{-1} \, V_L^\dagger \,.\nonumber
\end{eqnarray}
As stated in the introduction, here we assume the matrix $M_D$ to be known, namely the numerical values of the entries of the matrices $V_{L,R}$
and $M_D^{\rm diag}$ are given (see also appendix). Then it is clear form Eq.\,(\ref{eqMR}) that in order  to have a compact $M_R$ spectrum some particular conditions have to be assumed on the matrix $A$. Such conditions have been analyzed in \cite{Buccella:2012kc} (see section III.A of such a paper for a detailed discussion), and here we just report the main result. From \cite{Buccella:2012kc} one gets that a compact spectrum can be obtained if the matrix $A$ is such that
\begin{align} \label{weaker}
\abs{\frac{A_{33}}{A_{22}}} \leq \frac{M_{D_2}^2}{M_{D_3}^2}\,,\qquad \abs{\frac{A_{23}}{A_{22}}}\leq\frac{M_{D_2}}{M_{D_3}}\, .
\end{align}
\\In this way the entries of the matrix, which would be hierarchically large, are suppressed due to the largeness of $M_{D3}$. In the following for simplicity we assume an even more stringent condition by requiring
\begin{align} \label{stronger}
A_{23}=A_{33}=0 \, .
\end{align}
Under these assumptions, up to the first order in the small quantity $M_{D_1}^2 / M_{D_3}^2$, two of the three eigenvalues of the right-handed neutrino mass matrix are degenerate, in particular
\begin{eqnarray}
M_{R_1}&=& |A_{22}| M_{D_2}^2 \, , \nonumber\\
 M_{R_2}=M_{R_3}&=& |A_{13}| M_{D_1}M_{D_3} \, .
\end{eqnarray}
From  Eq.\,(\ref{eqMR}) it is clear that we can write the matrix $A$ in terms of observable neutrino mixing parameters as
\begin{eqnarray}\label{defA}
A&=& V_L^* \, U_{\textrm{PMNS}} \, (M_\nu^{\rm diag})^{-1}\, U_{\textrm{PMNS}}^T\, V_L^\dagger\,,
\end{eqnarray}
where
\begin{eqnarray} \label{pmns}
U_{\textrm{PMNS}}&=&\begin{pmatrix} 1 & 0 & 0 \\ 0 & \cos{\theta_{23}} & \sin{\theta_{23}} \\ 0 & -\sin{\theta_{23}} & \cos{\theta_{23}} \end{pmatrix} \begin{pmatrix} \cos{\theta_{13}} & 0 & \sin{\theta_{13}} e^{i\delta} \\ 0 & 1 & 0 \\ -\sin{\theta_{13}} e^{i\delta} & 0 & \cos{\theta_{13}} \end{pmatrix} \nonumber\\ 
&\times &\begin{pmatrix} \cos{\theta_{12}} & \sin{\theta_{12}} & 0 \\ -\sin{\theta_{12}} & \cos{\theta_{12}} & 0 \\ 0 & 0 & 1 \end{pmatrix} \begin{pmatrix} 1 & 0 & 0 \\ 0 & e^{i\alpha} & 0 \\ 0 & 0 & e^{i\beta} \end{pmatrix}
\end{eqnarray}
and $M_\nu^{\rm diag}$ is a function of the lightest active neutrino and of the two square mass differences $ \Delta m^2_{21}$ and $\Delta m^2_{31}$. In \eqref{pmns} $\theta_{12}$, $\theta_{23}$ and $\theta_{13}$ are the mixing angles, $\delta$ is the Dirac CP phase and $\alpha$ and $\beta$ are the Majorana phases. The two positions $A_{23}=A_{33} =0$ are complex equations that can be used to predict four neutrino mixing parameters from the other ones (remind that $V_L$ must be considered as given). Since the parameters $ m_1$,  $\delta$, $\alpha$, and $\beta$ are still experimentally undetermined\footnote{Here for simplicity we consider the normal hierarchy only. Note that $m_1$ has to be replaced by $m_3$ in case of inverse hierarchy.} the better choice is to use the two complex equations $A_{23}=A_{33} =0$ to obtain these unknown parameters as a function of the measured ones 
\begin{equation}
\Delta m^2_{21}, \, \Delta m^2_{31}, \, \theta_{12}, \, \theta_{23}, \, \theta_{13}  \,. \label{inpar}
\end{equation}
The latter ones are the input parameters of our numerical study that is shown in the Section 3.

\section{Neutrino mixing parameters and Leptogenesis}
It is convenient in this section to use a basis where the right-handed neutrino mass matrix is diagonal. In this basis for simplicity we denote
\begin{align}
\label{mRdiag}
M_R^{\rm diag}=W^{\dagger} M_R W^* = {\rm diag}(M_{1},M_{2},M_{3})\,.
\end{align}
where, since $M_R$ is symmetric, $W$ is a unitary matrix. The matrix $M_D$ then becomes
\begin{align}
\hat{M}_D=M_D W^* \,.
\end{align}
Let us define the $CP$ asymmetry in the decay of the $i$-th right-handed neutrino (with $i=1,2,3$) in the $l_\alpha$ lepton doublet (with $\alpha=1,2,3$) as the quantity
\begin{align}
\epsilon_{i\alpha}=\frac{\Gamma_{N_i\rightarrow l_{\alpha}\phi}-\Gamma_{N_i\rightarrow\overline{l}_{\alpha}\phi}}{\Gamma_{N_i\rightarrow l_{\alpha}\phi}+\Gamma_{N_i\rightarrow\overline{l}_{\alpha}\phi}}
\end{align}
whith $\Gamma$'s denoting the rates of the the corresponding decay processes. As well known, the asymmetries do not appear in a tree level computation of the decay rates, but rather they originate, at the lowest order, from the interference of tree level and one loop diagrams. The corresponding expressions, endowed with a regulating factor that gives contribution only in the case of quasi-degenerate neutrino mass spectrum\footnote{Although we have proven above that our case of study leads to a compact spectrum with two nearly degenerate neutrinos, it turns out that the numerical differences between the two masses, although extremely small, make the regulating factors negligible in determining the final value of the yield.}, are reported below
\begin{eqnarray}
\epsilon_{i\alpha}&=&\frac{1}{8\pi v^2} \sum_{k\not=i}\left[  A_{ik} \,f\left(\frac{M^2_k}{M^2_i}\right)
+B_{ik}\, g\left(\frac{M^2_k}{M^2_i}\right) \right]
\end{eqnarray}
where $v=174$ $GeV$ is the electroweak {\it v.e.v.}, $M_i$ are the masses defined in Eq. (\ref{mRdiag}) and
\begin{eqnarray}
f(x)&=&\sqrt{x}\left[\frac{1-x}{\left(1-x\right)^2 + \left( \frac{\Gamma_i}{M_i} - x\frac{\Gamma_k}{M_k}\right)^2} + 1 - \left(1+x\right) \log{\frac{1+x}{x}}\right]\,,\label{loop}\\
g(x)&=&\frac{1-x}{\left(1-x\right)^2 + \left( \frac{\Gamma_i}{M_i} - x\frac{\Gamma_k}{M_k}\right)^2}\, , \\
A_{ik}&=&\frac{\mbox{Im}\left[\hat{M}^\dagger_{Di\alpha} \hat{M}_{D\alpha k} \left( \hat{M}_D^\dagger \hat{M}_D \right)_{ik}\right]}{\left( \hat{M}_D^\dagger \hat{M}_D \right)_{ii}}\, , \\ B_{ik}&=&\frac{\mbox{Im}\left[\hat{M}^\dagger_{Di\alpha} \hat{M}_{D\alpha k} \left( \hat{M}_D^\dagger \hat{M}_D \right)_{ki}\right]}{\left( \hat{M}_D^\dagger \hat{M}_D \right)_{ii}}\,.
\end{eqnarray}
The total decay rate of the $i$-th right-handed neutrino, $\Gamma_i$, can be easily calculated from tree level diagrams as:
\begin{align}
\Gamma_i=\frac{M_i}{8\pi v^2}(\hat{M}_D^\dagger \hat{M}_D)_{ii} \, .
\end{align}
The evolution of right-handed neutrino species and the lepton asymmetries are described by a set of Boltzmann equations for the unknown abundances properly normalized, namely  $Y_X=n_X/ s$,  where $n_X$ is the number density of the $X$ species and $s=\frac{2\pi^2}{45} g^*_S T^3$ is the total entropy density\footnote{At the time of interest for leptogenesis $g^*_S=106.75$; this result derives from all Standard Model species being in equilibrium.}. In terms of the  abundance of left-handed $\alpha$-leptons the corresponding asymmetry is defined as $Y_{\Delta l_\alpha}\equiv Y_{l_\alpha}-Y_{\overline{l}_\alpha}$. Since sphaleronic processes, which are at equilibrium when leptogenesis occurs, preserve the charge $B-L$, it is convenient to express the equations in terms of the $B-L$ asymmetry for the $\alpha$-flavor $Y_{\Delta\alpha} \equiv Y_B/3-Y_{\Delta L_\alpha}$, where $Y_B$ is the total baryon asymmetry and $Y_{\Delta L_\alpha}$ is the total lepton asymmetry, involving both the left-handed and the right-handed leptons.  

The corresponding Boltzmann equations for right-handed neutrinos, written in terms of the dimensionless variable $z=\frac{M}{T}$ ($M$ being a convenient mass scale) involve only the term describing neutrino decays and inverse decays. The equations for the lepton asymmetries, instead, have to take into account neutrino decays and inverse decays, as well as the so-called \textit{washout processes}. These are all the processes (lepton and Higgs decays, inverse decays and scatterings) which tend to wash out the initial baryon asymmetry. The set equations then reads
\begin{eqnarray} \label{boltzmann}
sHz\frac{dY_i}{dz}&=&-\gamma_i \left(\frac{Y_i}{Y_i^{\rm eq}}-1\right)\label{neutrino}\\
sHz\frac{dY_{\Delta\alpha}}{dz}&=&-\sum_i \epsilon_{i\alpha}\gamma_i \left(\frac{Y_i}{Y_i^{\rm eq}}-1\right)+\sum_i \frac{\gamma_{i\alpha}}{2} \left(\frac{Y_{\Delta l_\alpha}}{Y_\alpha^{\rm eq}}+\frac{Y_{\Delta H}}{Y_H ^{\rm eq}}\right) \label{lepton}
\end{eqnarray}
where $i$ is the index for right-handed neutrinos. In these equations 
\begin{align}
\gamma_i =Y_i^{\rm eq} \, s \,  \Gamma_i \, \frac{K_1\left(\frac{M_i}{T}\right)}{K_2\left(\frac{M_i}{T}\right)}
\end{align}
is the thermally averaged total decay rate, with $K_{1,2}$ the first and second order modified Bessel functions, and $\gamma_{i\alpha}=\gamma_i P_{i\alpha}$ are the decay rates projected onto the $\alpha$ flavor\footnote{By this we mean the total decay rate in the $\alpha$ channel, comprised of both lepton and antilepton.}. The corresponding projectors are deduced, once again, by explicitly computing the tree level diagrams for the decay processes. The corresponding expressions are
\begin{align}
P_{i\alpha}=\frac{\hat{M}^\dagger_{Di\alpha} \hat{M}_{D\alpha i}}{\left( \hat{M}_D^\dagger \hat{M}_D\right)_{ii}}
\end{align}
Finally, the equilibrium abundances in \eqref{neutrino} and \eqref{lepton} are the following
\begin{itemize}
\item For right-handed neutrinos we used the equilibrium distribution function of a non relativistic particle with mass $M_i$ and $2$ degrees of freedom corresponding to the two polarizations of the Majorana neutrino $Y_i^{\rm eq}=\frac{45 M_i^2 z^2}{2\pi^4 g^*_S M^2} K_2\left(\frac{M_i z}{M}\right)$;
\item For the Higgs particle we used the equilibrium distribution function of a massless boson with the $2$ degrees of freedom corresponding to the $SU(2)$ doublet structure of the Higgs $Y_H^{\rm eq}=\frac{45\zeta(3)}{g^*_S\pi^4}$.
\item For the lepton doublet we used the equilibrium distribution function of massless fermions with the $2$ degrees of freedom corresponding to the $SU(2)$ doublet structure of the particles $Y_\alpha^{\rm eq}=\frac{135\zeta(3)}{4\pi^4g^*_S}$.
\end{itemize}

In order to have the Boltzmann equations  only in terms of $Y_i$ and $Y_{\Delta\alpha}$ it is necessary to 
 express all asymmetries in the form $Y_{\Delta l_\alpha}=A_{\alpha\beta}Y_{\Delta\beta}$ and $Y_{\Delta H}=C_{\beta}Y_{\Delta\beta}$, where \cite{Abada:2006ea}
\begin{align}
A=\frac{1}{2148}\begin{pmatrix} -906 && 120 && 120\\ 75 && -688 && 28\\ 75 && 28 && -688 \end{pmatrix}, ~~~~~ C=-\frac{1}{358} \begin{pmatrix} 37 \\ 52 \\ 52 \end{pmatrix}\,.
\end{align}
These relations are deduced from the equilibrium conditions for the reactions occurring at the time of leptogenesis. We stress here that, since reactions can go out of equilibrium at specific temperatures, they strongly depend on the value of temperature when leptogenesis occurs.

The set of equations \eqref{boltzmann} can now be numerically solved to obtain the asymptotic values of $Y_{\Delta\alpha}$. Once this has been done, the asymptotic value of the baryon asymmetry yield can be found through the sphaleron relation, derived from imposing the equilibrium condition on the chemical potential\footnote{A subtlety lies in the fact that these equilibrium conditions must be imposed not at the time at which leptogenesis happens; in fact, lepton number is produced at this time, but continues to be converted to baryon number until sphalerons run out of equilibrium. It is at this temperature that the equilibrium conditions must be imposed; therefore, the numerical coefficient in \eqref{sphaleron} does not depend on the leptogenesis temperature.}:
\begin{align} \label{sphaleron}
Y_{\Delta B}=\frac{28}{79}\sum_\alpha Y_{\Delta\alpha}\,.
\end{align}
The experimental value of the baryon asymmetry yield is given by \cite{Ade:2015xua}
\begin{align} \label{exp}
Y_{\Delta B}=(8.65\pm 0.06)\cdot 10^{-11}{\rm ~~~~(68\%~ C.L.)}\,.
\end{align}

\section{Results}

In order to proceed numerically we have to fix the Dirac neutrino mass matrix, $m_D$. Even if the procedure deligned in the previous section is completely general here we consider a $SO(10)$  inspired model that is described in more detail in appendix. In the $SO(10)$ framework we expect  
\begin{eqnarray}
&& M_D\approx M_{\textrm{up}}\,; \qquad M_{\ell}\approx M_{\textrm{down}}
\end{eqnarray}
Assuming the down quark $M_{\textrm{down}}$ and charged lepton $M_{\ell}$ mass matrices approximatively diagonal, it follows that the up quark mass matrix must be diagonalized by the CKM mixing matrix. Moreover if the scalar sector of the $SO(10)$
model does not contain the {\bf 120 } irreducible representation, then both $m_D$ and $ M^{\textrm{up}}$
are symmetric and 
\begin{eqnarray}\label{mDs}
&& M_D\approx V_{\textrm{CKM}}^\dagger M_{\textrm{up}}^{\rm diag} V_{\textrm{CKM}}^*
\end{eqnarray}
where $M_{\textrm{up}}^{\rm diag} \equiv \{ m_{u},m_c,m_t \}$ that are the physical up, charm and top quark masses
and 
\begin{eqnarray}
V_{\textrm{CKM}}=&&\begin{pmatrix} 1 & 0 & 0 \\ 0 & \cos{\omega_{23}} & \sin{\omega_{23}} \\ 0 & -\sin{\omega_{23}} & \cos{\omega_{23}} \end{pmatrix} \begin{pmatrix} \cos{\omega_{13}} & 0 & \sin{\omega_{13}} e^{i\omega} \\ 0 & 1 & 0 \\ -\sin{\omega_{13}} e^{i\omega} & 0 & \cos{\omega_{13}} \end{pmatrix} \nonumber\\
&&  \,\begin{pmatrix} \cos{\omega_{12}} & \sin{\omega_{12}} & 0 \\ -\sin{\omega_{12}} & \cos{\omega_{12}} & 0 \\ 0 & 0 & 1 \end{pmatrix}\,,
\end{eqnarray}
where the values used for angles, masses and the phase are reported in Table\,(\ref{benchmark}). From Eq.\,(\ref{mDs}) one gets that $V_L$ is fixed, namely $V_L\equiv V_{\textrm{CKM}}$, and it can be replaced in Eq.\,(\ref{defA}).  As already stated before, by using the conditions  $A_{23}=A_{33} =0$ in Eq.\,(\ref{defA}) one gets $m_1,\, \delta,\, \alpha,\, \beta$ as a function of the input parameters (\ref{inpar}).

The current bounds on neutrino mixing parameters can be found for example in Ref.s\,\cite{Esteban:2016qun,Capozzi:2017ipn,deSalas:2017kay}. For the present analysis we refer to the values given in \cite{Capozzi:2017ipn} \footnote{The results of References \cite{Capozzi:2017ipn,deSalas:2017kay} are consistent within $1\sigma$, while the value of the atmospheric mixing angle reported in the recent update NuFIT 3.2 (2018), {\it www.nu-fit.org}, differs from the other two analyses for more than $1\sigma$.}.  In order to simplify the analysis we fix $\Delta m^2_{21}$ and $\Delta m^2_{31}$ to their best fit  values, and we take randomly the three mixing angles within the corresponding $3\sigma$ ranges. In particular we randomly extract 10000 points in this three dimensional space. For each input random point  (\ref{inpar}) we get a set of output values for the lightest neutrino mass and the three neutrino phases. 

\begin{figure}[h!]
\begin{center}
\includegraphics[width=0.42\textwidth]{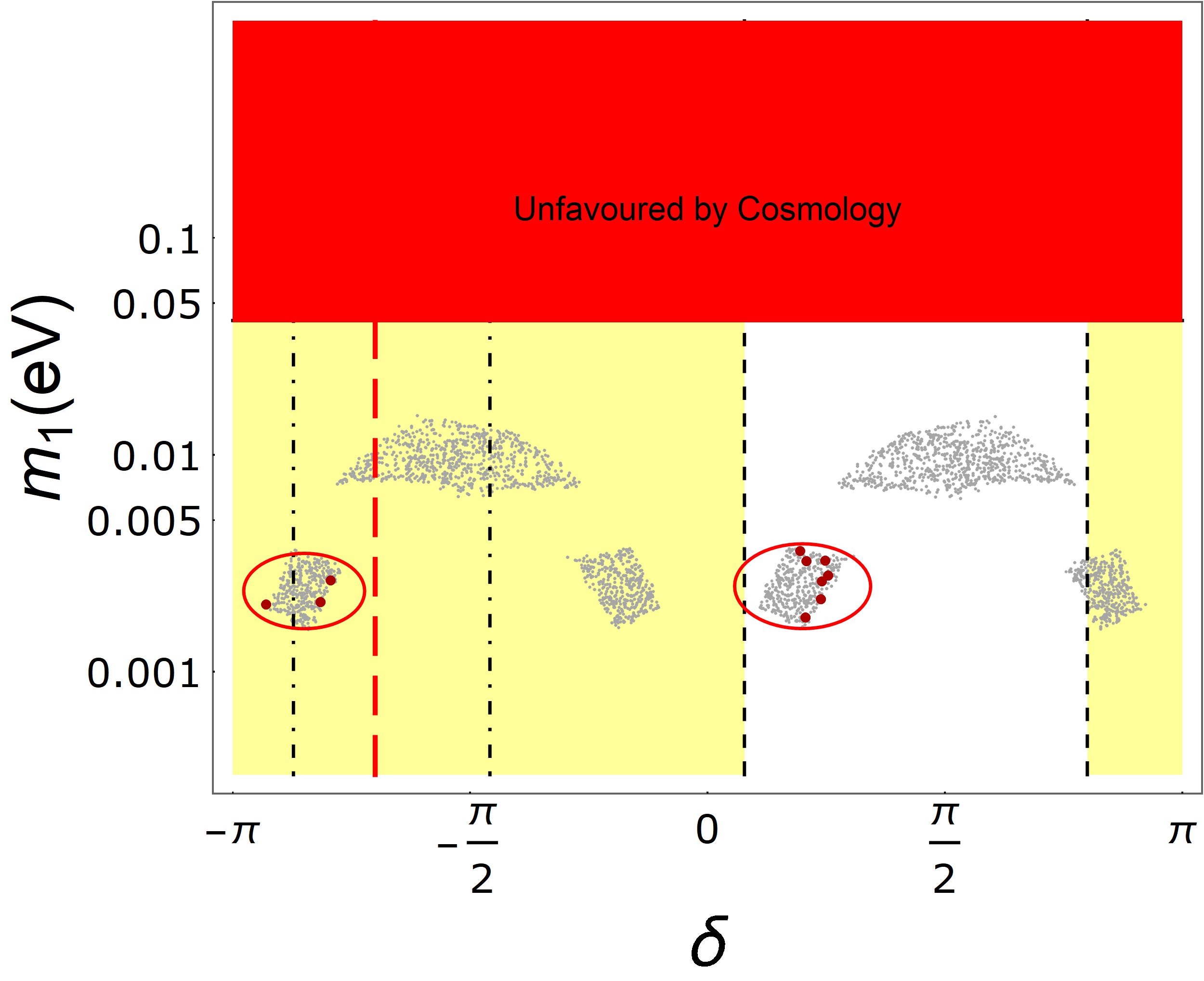}
\includegraphics[width=0.40\textwidth]{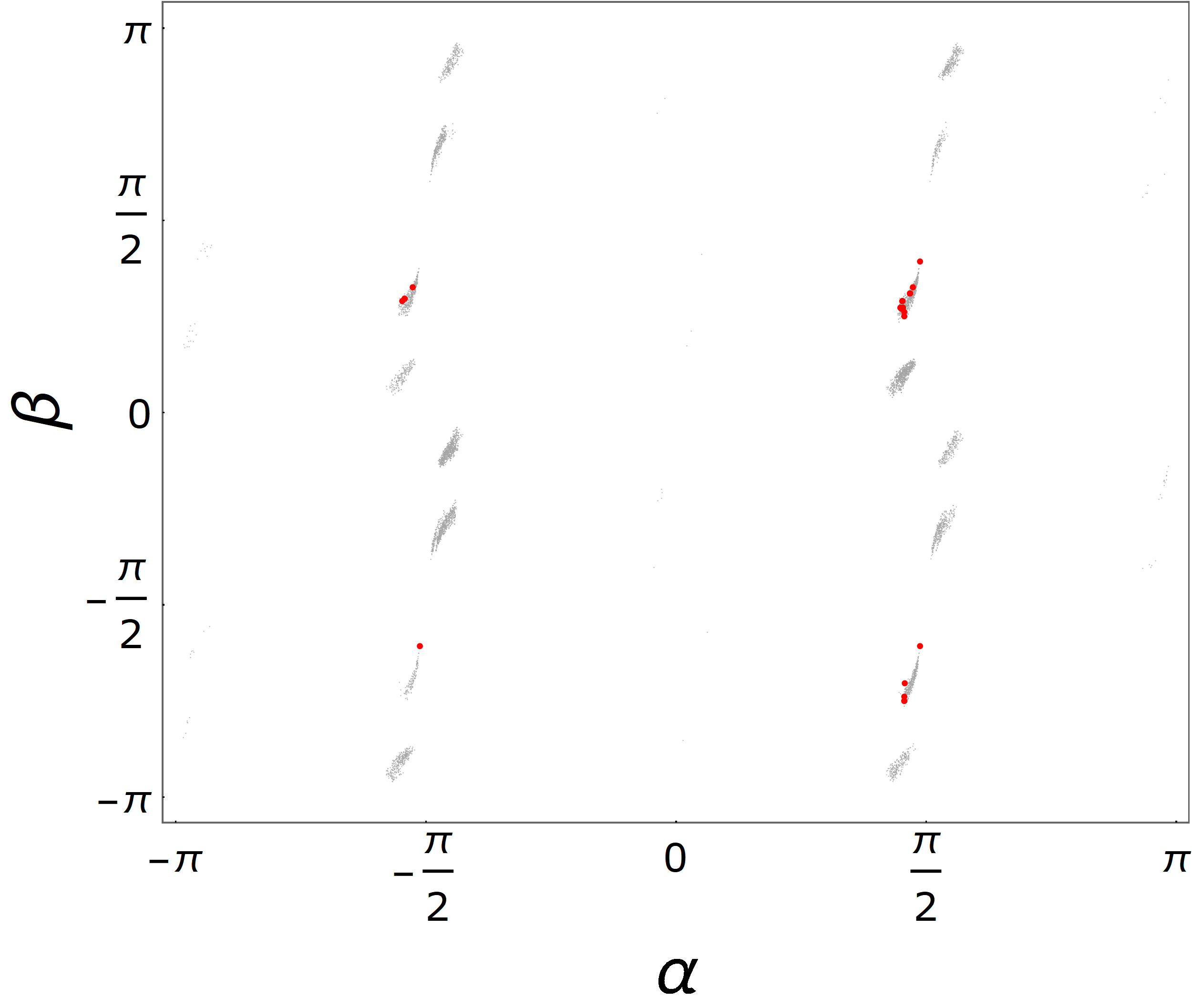}
\end{center}
\caption{\label{fig:1} (Left panel) Lightest neutrino mass vs the Dirac phase. The different vertical bands correspond to the experimental values for the Dirac phase (see text for details) and the horizontal band is the upper limit coming from Cosmology. In yellow the $3\sigma$ confidence band is evidenced. (Right panel) Majorana phases $\alpha$ and $\beta$ for the numerically generated points. In both plots red points are obtained by imposing $Y_{\Delta B}$ within the 3-$\sigma$ experimental range while grey points are not constrained from baryon asymmetry.}
\end{figure}

The results are given in Figure \ref{fig:1} where we show the mass $m_1$ as a function of the Dirac phase $\delta$ (left panel) and the correlation between the Majorana phases $\alpha$ and $\beta$ (right panel). In the left plot we report the $1\sigma$ (dot-dashed lines) and the $3\sigma$ (dashed lines) experimental range for the Dirac phase, while the dashed red line is the best fit value of the Dirac phase. We note that under the above hypotheses (see Eq. \ref{stronger}) a link between the lightest neutrino mass and the Dirac phase is expected. The scatter plot of Figure \ref{fig:1} (grey points in left panel) shows such correlation when the uncertainty on the input parameters is taken into account. It is interesting to observe that such a correlation is minimally smeared out by these observational uncertainties. Despite the high number of 10000 generated points, only a small part of these (about 30) led to a value for the baryon asymmetry consistent with the experimental value given in \eqref{exp} within a $3\sigma$ range. For the benefit of the reader we highlight the regions containing these points by circling them in red. We notice that one of these regions falls outside the $3\sigma$ confidence range for $\delta$.
Moreover, the lightest neutrino mass has an upper and lower limit that is not affected by present cosmological observations; the regions efficient for leptogenesis, in fact, have $m_1\sim (0.002,0.004)$ $eV$ and $\delta\sim(-0.90\pi,-0.75\pi)$. Also the Majorana phases are constrained, in particular $|\alpha|\approx\pi/2$ and $\beta$ is distributed in the range $(-\pi,+\pi)$ in two isolated regions centered around the values of $0.3\pi$ and $-0.7\pi$.

The neutrinoless double beta parameter $m_{\beta\beta}=\abs{\sum_i m_i U^2_{PMNS_{ei}}}$ is found to lie between $0.001$ $eV$ and $0.02$ $eV$, which is below the experimental bounds, set, for example, in \cite{Agostini:2018tnm}\footnote{For a comparison with major neutrinoless double beta experiments see Table 2 of \cite{Agostini:2018tnm}.}.

\begin{figure}[h!]
\begin{center}
\includegraphics[width=0.42\textwidth]{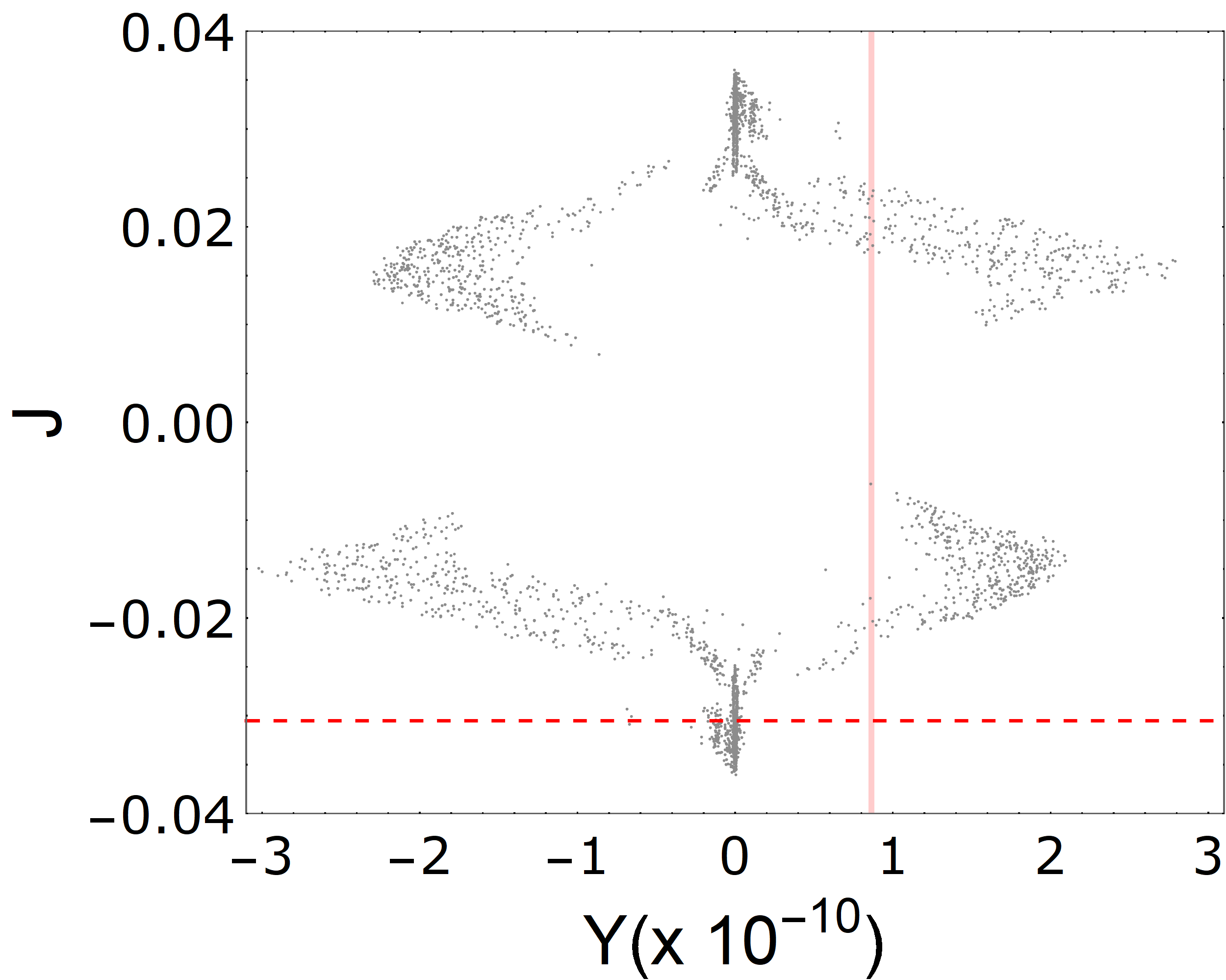}
\caption{Jarlskog invariant parameter as a function of the baryon asymmetry yield obtained. The vertical band correspond to the $3-\sigma$ experimental value The horizontal dashed line represents the value obtained by fixing all the oscillating parameters to their best fit values.}\label{figmix}
\end{center}
\end{figure}
In Figure\,\ref{figmix} we give the Jarlskog parameter \cite{Jarlskog:1985ht} as a function of the baryon asymmetry yield, together with the red vertical band representing the $3\sigma$ confidence range for the baryon abundance experimentally measured. We see that requiring a baryon asymmetry within about a $3-\sigma$ range around the experimental value of the baryon abundance, we get $J$ in the range $(-0.022,-0.018)$ (approximately independent of the sign). It is to be noted, however, that different signs predict different yields, due to the opposite value of the CP asymmetry.

\section{Conclusions}
In this paper we analize the baryogenesis \textit{via} leptogenesis scenario for a type-I seesaw mechanism at the basis of neutrino mass generation. In this framework, we assume a reasonable structure for Dirac neutrino mass matrix, namely symmetric and similar to up-type quark mass matrix, like occurring in SO(10) inspired models. These assumptions imply a relevant correlation between the parameters of CP violation at high and low energy, and this restricts  low energy neutrino parameter space (already compatible with neutrino phenomenology) once we require consistency with the observed baryon asymmetry.
Unfortunately, with the hierarchical structure induced on right-handed neutrino mass matrix by previous similarity hypothesis, it is not possible to obtain a 
viable leptogenesis realization, because the lightest right-handed neutrino mass results to be below the Davidson-Ibarra limit. One can circumvent this problem by imposing a fine tuning in the neutrino parameters, which providing a compact right-handed neutrino spectrum,  allows to obtain an efficient leptogenesis. This fine tuning, through the numerical resolution of the Boltzmann equations ruling the yields evolution, provides the following allowed intervals for the lightest neutrino mass and the Dirac CP phase (in a $3 \sigma$ range from the experimental values): $-0.90\pi<\delta<-0.75\pi$ and   $m_1\sim( 0.002 - 0.004)$\,eV. 

\section*{Acknowledgements}
We thanks Marco Chianese for useful discussion and for providing us a benchmark point for testing our numerical solution of the Boltzman equation for leptogenesis.

\vskip10.mm

\section*{Appendix}
\appendix
\section{Examples of hierarchical lepton Yukawas}

\vskip4.mm
\noindent{\bf First example: FN scenarios}\\
Assuming FN families symmetries we expect quite hierarchical Dirac Neutrino Yukawa interactions. 
Here we  provide an example from reference \cite{Altarelli:1999wi} where the three left-handed $SU(2)_L$ doublets have $U(1)_{FN}$ charges $(3,0,0)$, the charged lepton right-handed $(3,2,0)$ and the right-handed singlets $(1,-1,0)$. Moreover two FN scalars electroweak singlets has been introduced, $\theta$ and $\theta'$, with charges 1 and $-1$ respectively. Under such assumptions the resulting mass matrices are proportional to
\begin{equation}
M^\ell\propto 
\left(
\begin{array}{ccc}
\lambda^6 & \lambda^5&\lambda^3\\
\lambda^5 & \lambda^4&\lambda^2\\
\lambda^3 & \lambda^2&1
\end{array}
\right)\,,\quad
m_D\propto 
\left(
\begin{array}{ccc}
\lambda^2 & 1&\lambda\\
1 & \lambda'^2&\lambda'\\
\lambda & \lambda'&1
\end{array}
\right)\,,\quad
M_R\propto 
\left(
\begin{array}{ccc}
\lambda^2 & 1&\lambda\\
1 & \lambda'^2&\lambda'\\
\lambda & \lambda'&1
\end{array}
\right)\,,
\end{equation}
where $\lambda\propto \langle \theta\rangle$ and $\lambda'\propto \langle \theta'\rangle$. The proportional symbol is just to remember that each entry is multiplied by an arbitrary  order one parameter. From such matrices we obtain
\begin{equation}
m_\nu\propto 
\left(
\begin{array}{ccc}
\lambda^6 & \lambda^3&\lambda^3\\
\lambda^3 & 1&1\\
\lambda^3 & 1&1
\end{array}
\right)
\end{equation}
predicting a large atmospheric angle. However, here we are not interested to give a realistic model that can explain all fermion observables but we want just to show an example of a Dirac Yukawa coupling providing hierarchical eigenvalues. Indeed, fixing $\lambda=0.05$ and $\lambda'=0.1$ we obtain eigenvalues of order $\mathcal{O}(10^{-9},10^{-3},1)$.

\vskip4.mm
\noindent{\bf Second example: $SO(10)$ scenarios}\\
In minimal $SO(10)$ models it is well known that the up-type quark mass matrix and the Dirac neutrino one are approximatly equal $M_{up}\approx m_D$. We note that if only one {\bf 10} irreducible representation appears in the scalar sector, then $M_{up}= m_D$ exactly. On the other hand, in this case the CKM would be exactly the identity matrix, leading to an unrealistic model. A possible minimal solution is to add a ${\bf \overline{126}}$ scalar representation \cite{Babu:1992ia}. The fit of all Standard Model observables in this minimal scenario has been studied in detail for instance in Ref. \cite{Bertolini:2006pe}. 
We also observe that if the scalar sector of the theory does not contain the  {\bf 120} irreducible representation of $SO(10)$, the two matrices $M_{up}$ and $ m_D$ are symmetric.
These facts imply that $m_D$ is diagonalized by a single unitary matrix (and not two) with quite small angles of the order of the CKM mixing matrix\cite{Patrignani:2016xqp}, and that its eigenvalues are hierarchical like the up quark masses. 

We moreover consider an extension of the minimal $SO(10)$ scalar spectrum (with only {\bf 10} and  ${\bf \overline{126}}$ scalar representations), with a {\bf 45} multiplet that takes v.e.v. in the $T_{3R}$ direction of $SO(10)$ \cite{Anderson:1993fe}. In this case the non-renormalizable operator ${\bf 16\,16\,45\, \overline{126}}$ gives a contribution only to right handed neutrino. Therefore, differently from the minimal $SO(10)$ model, the right handed neutrino mass is a free matrix.

\section{Benchmark point}
In this appendix we explicitly report a point obtained in our numerical analysis which can be used as benchmark or to test our results; in Table \ref{benchmark} we report the values of the parameters from the Standard Model (where $\omega$ has been used to denote the mixing angles and the $CP$ phase of the CKM matrix, in order to avoid confusion with the angles of the PMNS matrix), the values of the parameters generated for the specific point given and the equilibrium yield obtained.
\begin{table}[h!]
\begin{center} \label{benchmark}
\begin{tabular}{|l|l|l|}
\hline
\multirow{12}*{Constants} & $G_F$ $(GeV^{-2})$ & $1.166\times10^{-5}$ \\
\cline{2-3}
& $M_W$ $(GeV)$ & $80.39$\\
\cline{2-3}
& $v(GeV)$ & $174$\\
\cline{2-3}
& $m_u$ $(eV)$ & $6.7\times10^{5}$\\
\cline{2-3}
& $m_c$ $(eV)$ & $0.327\times10^{9}$\\
\cline{2-3}
& $m_t$ $(eV)$ & $99.1\times10^9$\\
\cline{2-3}
& $\Delta m^2$ $(eV^2)$ & $3.84\times10^{-3}$\\
\cline{2-3}
& $\delta m^2$ $(eV^2)$ & $11.8\times10^{-5}$\\
\cline{2-3}
& $\omega_{12}$ $(^\circ )$ & $13.02$ \\
\cline{2-3}
& $\omega_{13}$ $(^\circ)$ & $0.20$\\
\cline{2-3}
& $\omega_{23}$ $(^\circ)$ & $2.35$\\
\cline{2-3}
& $\omega$ & $1.20$\\
\hline
\multirow{3}*{Input parameters} & $\theta_{12}$ & $-0.580$ \\
\cline{2-3} 
& $\theta_{13}$ & $0.142$ \\
\cline{2-3} 
& $\theta_{23}$ & $0.861$ \\ 
\hline
\multirow{5}*{Output results} & $m_1$ $(eV)$ & $0.0026$ \\
\cline{2-3} 
& $\delta$ & $3.505$ \\
\cline{2-3} 
& $\alpha$ & $1.512$ \\
\cline{2-3} 
& $\beta$ & $-2.052$ \\
\cline{2-3}
& $Y$ & $1.28\times 10^{-10}$\\
\hline
\end{tabular}
\end{center}\caption{Input parameters used in our numerical analysis and corresponding output for a benchmark point.}\label{benchmark}
\end{table}
We also report the $CP$ asymmetry matrix for the above mentioned point:
\begin{align}
\epsilon=\begin{pmatrix} 4.43\times10^{-7} & 2.54\times10^{-6} & -2.96\times10^{-6}\\
5.03\times10^{-8}&1.09\times10^{-6}&6.72\times10^{-4}\\
4.98\times10^{-8}&1.09\times10^{-6}&6.72\times10^{-4}
\end{pmatrix}
\end{align}

\end{document}